\documentclass[twocolumn,noshowpacs,preprintnumbers,amsmath,amssymb]{revtex4}

\usepackage{graphicx}% Include figure files
\usepackage{dcolumn}% Align table columns on decimal point
\usepackage{bm}% bold math
\usepackage{times,mathptmx}
\usepackage{color}
\usepackage{hyperref}

\begin{document}

\title{Quasi-two-dimensional optomechanical crystals with a complete phononic bandgap}

\author{Thiago P. Mayer Alegre} \altaffiliation{These two authors contributed equally.}
\author{Amir Safavi-Naeini} \altaffiliation{These two authors contributed equally.}
% Lines break automatically or can be forced with \\
\author{Martin Winger}
\author{Oskar Painter}
\email{opainter@caltech.edu}
\homepage{http://copilot.caltech.edu}
\affiliation{%
Thomas J. Watson, Sr., Laboratory of Applied Physics, California Institute of Technology, Pasadena, CA 91125}%

\date{\today}

\begin{abstract}
A fully planar two-dimensional optomechanical crystal formed in a silicon microchip is used to create a structure devoid of phonons in the GHz frequency range.  A nanoscale photonic crystal cavity is placed inside the phononic bandgap crystal in order to probe the properties of the localized acoustic modes.  By studying the trends in mechanical damping, mode density, and optomechanical coupling strength of the acoustic resonances over an array of structures with varying geometric properties, clear evidence of a complete phononic bandgap is shown.  
\end{abstract}

\maketitle

Interest in cavity-optomechanical systems, in which light is used to sensitively measure and manipulate the motion of a mechanically compliant optical cavity, has grown rapidly in the last few years due to the demonstration of micro- and nano-scale systems in which the radiation pressure force of light is manifest.  The physics of these systems is similar in many regards to the inelastic scattering of light by localized molecular vibrations (Raman scattering), where the mechanical resonance is now associated with a moveable end-mirror affixed to a spring (or hung as a pendulum), as is the case in a Fabry-Perot cavity.  More complex cavity-optomechanical geometries, such as whispering-gallery mode structures~\cite{Kippenberg2005,Carmon2005,qiang_lin_coherent_2010,rosenberg_static_2009,wiederhecker_controlling_2009,Sridaran2010}, nanomembranes placed within Fabry-Perot cavities, and near-field optical and microwave devices utilizing the gradient force~\cite{eichenfield_picogram-_2009,Regal08}, have also been demonstrated to produce strong radiation pressure effects.  Two structures with particular engineerability of optical and mechanical properties are the photonic crystal fiber (PCF)~\cite{dainese_stimulated_2006,wiederhecker_coherent_2008} and the recently demonstrated chip-scale optomechanical crystal (OMC)~\cite{eichenfield_optomechanical_2009}. In the case of the OMC cavity, a combined photonic and phononic crystal formed in a thin nanobeam of silicon (Si) is used to create a $200$~THz optical cavity simultaneous with a GHz acoustic cavity. 

%%%%%%%%%%%%%%%%%%%%%%%%%%%%%%%%%%% FIGURE 01 %%%%%%%%%%%%%%%%%%%%%%%%%%%%%%%%%%%%%%%%%%%%%%%%%%%%%%%%
\begin{figure}[ht!]
\includegraphics[width=0.95\columnwidth]{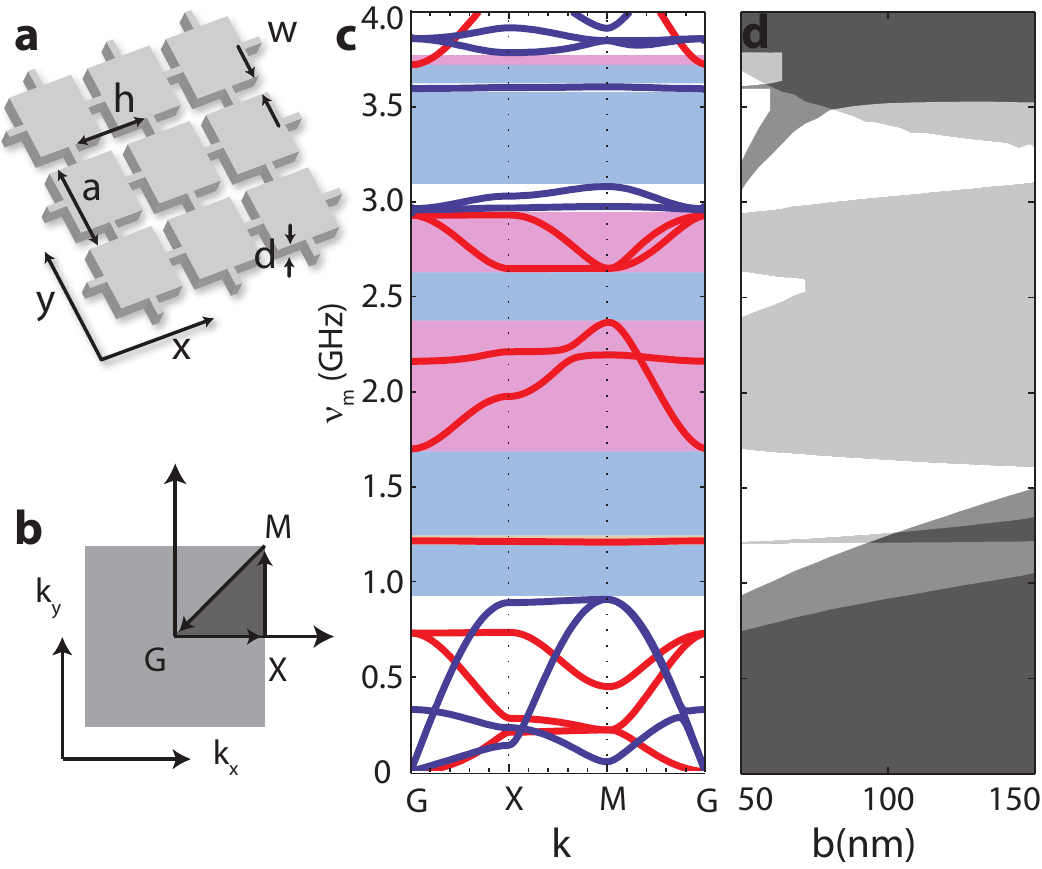} %single column figure - for two-column format
\caption{\label{Fig1} \textbf{a}, Real space crystal lattice of the cross crystal with lattice constant $a$, cross length $h$, cross width $w$, and membrane thickness $d$. The bridge width is defined as $b=a-h$. \textbf{b}, Reciprocal lattice of the first Brillouin zone for the cross crystal. \textbf{c}, Phononic band diagram for the nominal cross structure with $a=1.265~\mu$m, $h=1.220~\mu$m, $w=340~$nm. Dark blue lines represent the bands with even vector symmetry for reflections about the $x-y$ plane, while the red lines are the flexural modes which have odd vector mirror symmetry about the $x-y$ plane. \textbf{d}, Tuning of the bandgap with bridge width, $b$. Light grey, dark grey, and white areas indicate regions of a symmetry-dependent (i.e., for modes of only one symmetry) bandgap, no bandgap, and full bandgap for all acoustic modes, respectively.}
\end{figure}
%%%%%%%%%%%%%%%%%%%%%%%%%%%%%%%%%%%%%%% FIGURE 01 %%%%%%%%%%%%%%%%%%%%%%%%%%%%%%%%%%%%%%%%%%%%%%%%%%%%%%%%

%%%%%%%%%%%%%%%%%%%%%%%%%%%%%%%%%%%%%% FIGURE 02 %%%%%%%%%%%%%%%%%%%%%%%%%%%%%%%%%%%%%%%%%%%%%%%%%%%%%%%%
\begin{figure*}[ht!]
\includegraphics[width=1.95\columnwidth]{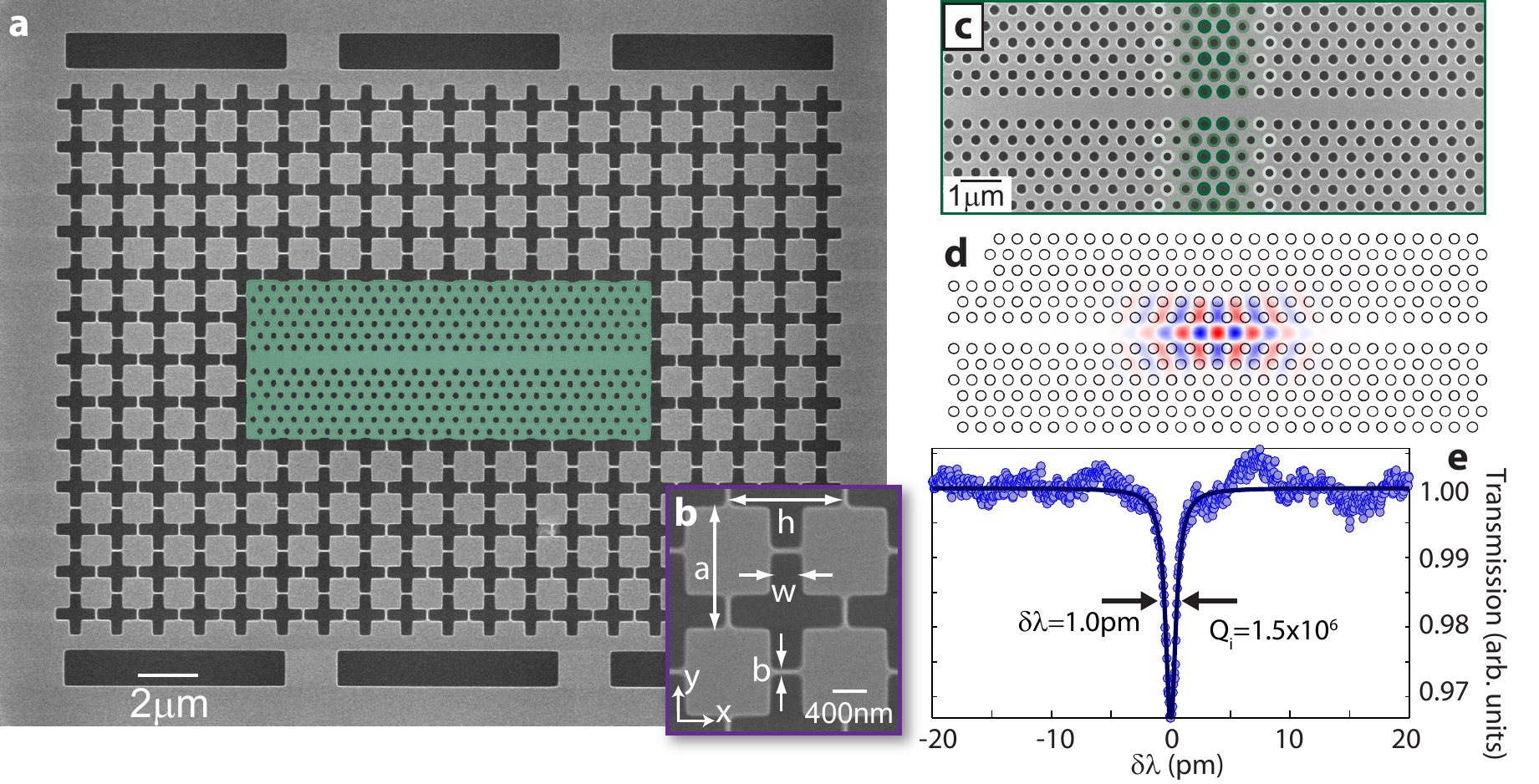} %double column figure - for two-column format
\caption{\label{Fig2} \textbf{a}, Scanning electron micrograph (SEM) of one of the fabricated 2D-OMC structures. The photonic nanocavity region is shown in false green color. In \textbf{b}, Zoom-in SEM image of the cross crystal phononic bandgap structure. \textbf{c}, Zoom-in SEM image of the optical nanocavitywithin embedded in the phononic bandgap crystal. Darker (lighter) false colors represents larger (smaller) lattice constant in the optical cavity defect region. \textbf{d}, FEM simulation of $E_\text{y}$ electrical field for the optical cavity. \textbf{e}, Typical measured transmission spectra for the optical nanocavity, showing a bare optical $Q$-factor of $Q_i=1.5\times10^6$.}
\end{figure*}
%%%%%%%%%%%%%%%%%%%%%%%%%%%%%%%%%%%%%%%%%%%% FIGURE 02 %%%%%%%%%%%%%%%%%%%%%%%%%%%%%%%%%%%%%%%%%%%%%%%%%%%%%%%%

%%%%%%%%%%%%%%%%%%%%%%%%%%%%%%%%%%%%%%%%%%% FIGURE 03 %%%%%%%%%%%%%%%%%%%%%%%%%%%%%%%%%%%%%%%%%%%%%%%%%%%%%%%%
\begin{figure}[ht!]
\includegraphics[width=\columnwidth]{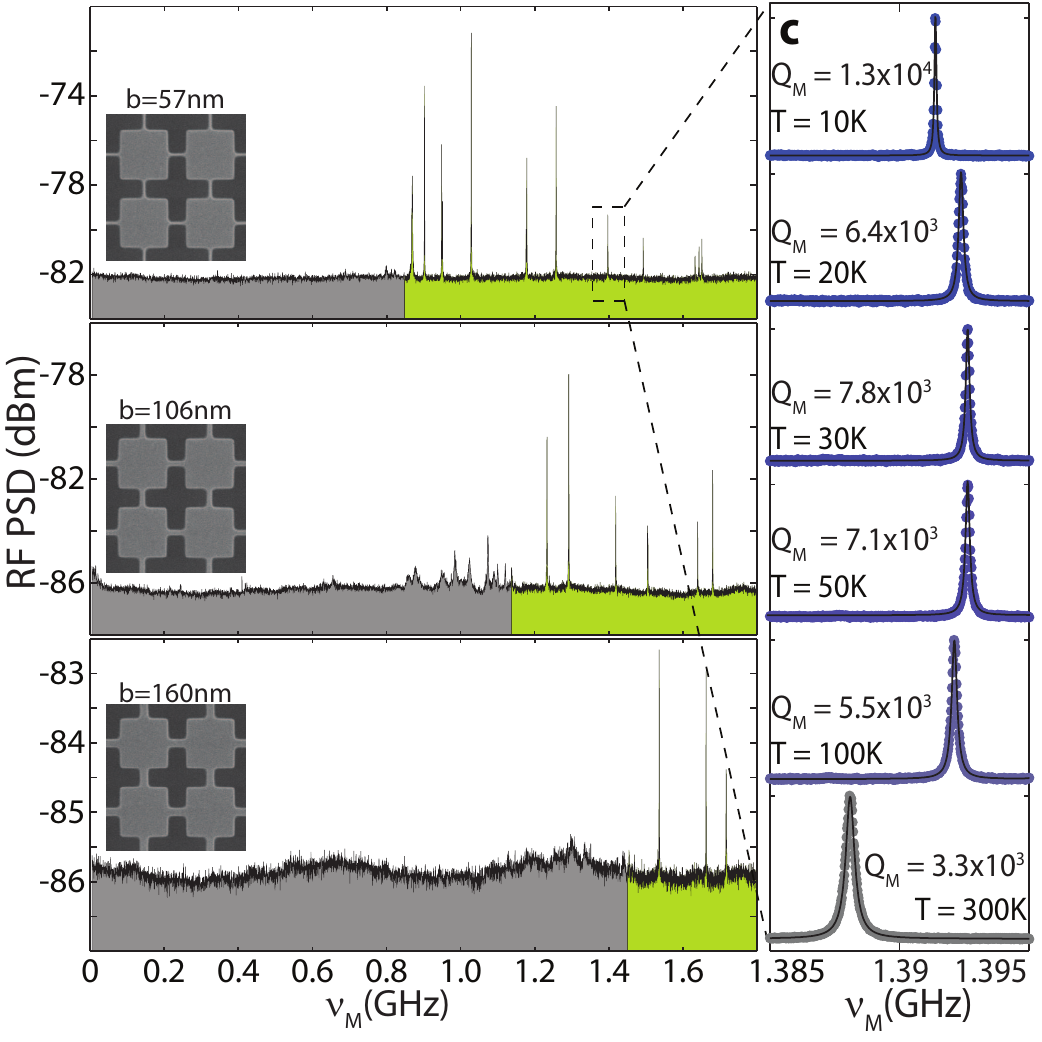} %single column figure  - for preprint format
\caption{\label{Fig3} \textbf{a}, Optically transduced RF power spectral density of the thermal Brownian motion of $S_{1}$ structures with $b=57$~nm (top), $b=106$~nm (middle), and $b=160$~nm (bottom).  The inferred region below the phononic bandgap is shaded grey (see main text).  \textbf{b}, Temperature dependence of the mechanical quality factor for the $1.4$~GHz acoustic mode of the $S_{1}$ structure with $b=57$~nm.}
\end{figure}
%%%%%%%%%%%%%%%%%%%%%%%%%%%%%%%%%%%%%%%%%%%% FIGURE 03 %%%%%%%%%%%%%%%%%%%%%%%%%%%%%%%%%%%%%%%%%%%%%%%%%%%%%%%%

%%%%%%%%%%%%%%%%%%%%%%%%%%%%%%%%%%%%%%%%%%%% FIGURE 04 %%%%%%%%%%%%%%%%%%%%%%%%%%%%%%%%%%%%%%%%%%%%%%%%%%%%%%%%
\begin{figure*}[ht!]
\includegraphics[width=2\columnwidth]{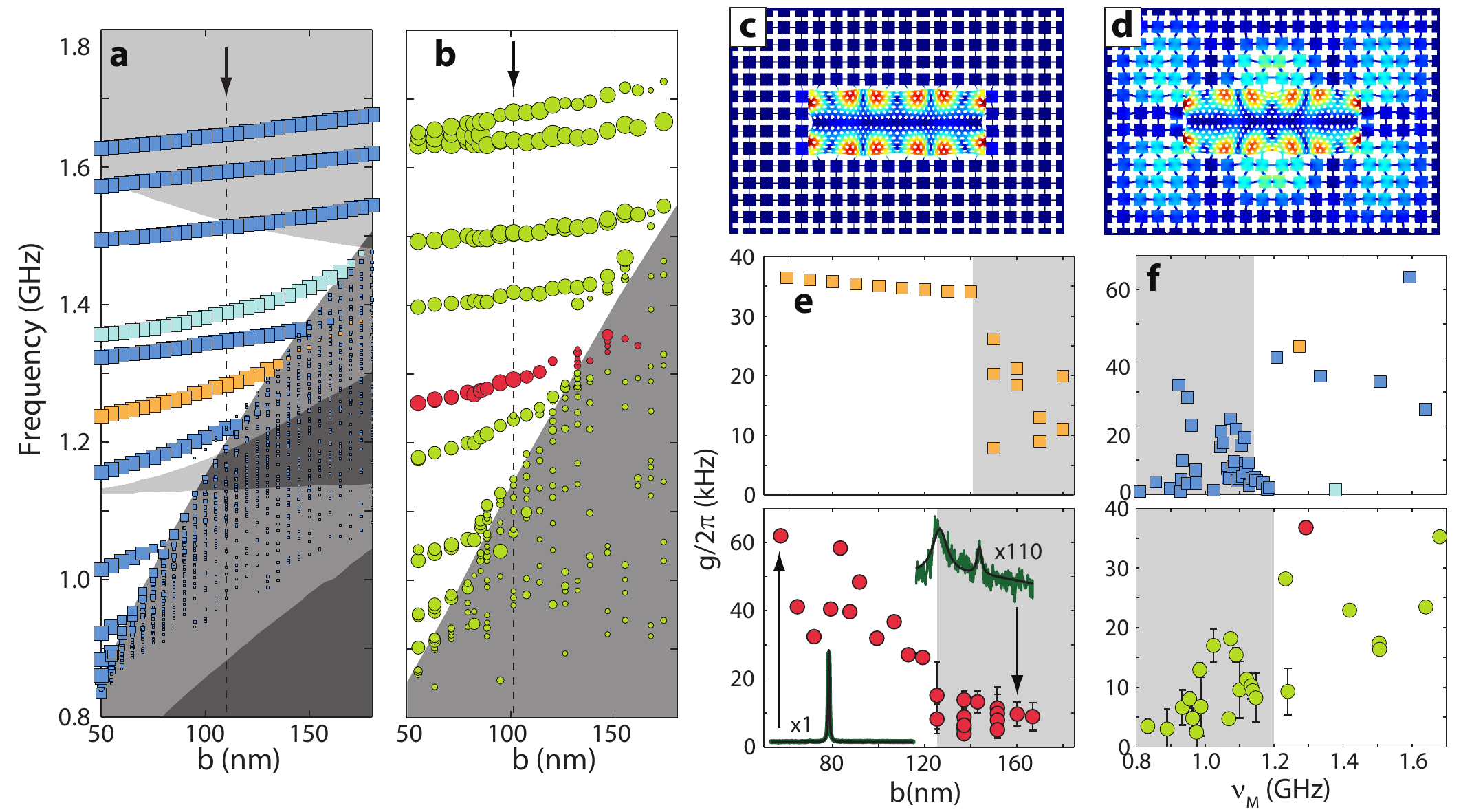} %double column figure - for two-column format
\caption{\label{Fig4} \textbf{a}, Plot of the 3D-FEM simulated in-plane localized acoustic modes of the $S_{1}$ structure as a function of bridge width $b$. Each marker corresponds to a single acoustic mode, with the marker size proportional to the logarithm of the calculated acoustic radiation $Q$-factor. The light blue shaded markers correspond to acoustic bands which are optically dark. The shading corresponds to the same color coding of the phononic bandgaps as that used in Fig.~\ref{Fig1}d.  \textbf{b}, Measured mode plot of the optically-transduced localized acoustic modes for an array of $S_1$ structures with varying bridge width.  The marker size of each resonance is related to the logarithm of the measured mechanical $Q$-factor. The inferred spectral region below the phononic bandgap is shaded grey.  \textbf{c} and \textbf{d}, FEM simulations of the displacement field amplitude ($|\mathbf Q(\mathbf{\text{r}})|$) for the acoustic mode in the orange colored band around $1.35$~GHz in \textbf{a}. In \textbf{c} the mode is within the phononic band gap resulting in a radiation-limited $Q_\text{M}^\text{(rad)}\approx10^9$. In \textbf{d} the mode is on the edge of the bandgap and has a reduced $Q_\text{M}^\text{(rad)} < 10^3$. \textbf{e}, Simulated ($\square$) and measured ($\circ$) optomechanical coupling rate $g$ for the orange (red) highlighted acoustic band in \textbf{a} (\textbf{b}). \textbf{f}, Simulated ($\square$) and measured ($\circ$) optomechanical coupling rate $g$ for the series of acoustic modes of the $S_{1}$ structure with $b\sim 100$~nm (vertical dashed curves in \textbf{a} and \textbf{b})}
\end{figure*}
%%%%%%%%%%%%%%%%%%%%%%%%%%%%%%%%%%%%%%%%%%% FIGURE 04 %%%%%%%%%%%%%%%%%%%%%%%%%%%%%%%%%%%%%%%%%%%%%%%%%%%%%%%%

%Despite some similarities, phononic and photonic devices work in very different frequency (energy) ranges. The low frequency range provided by hypersonic devices make them more susceptible to thermal noise when compared to optical system. However they are the natural choice for coherently interaction with radio frequency (RF) electromagnetic waves. While communication system rely on efficient transmission over long distances and large bandwidth (~THz) provided by optical system, narrow band filters encounter their optimal design in bulk-acoustic-waves (BAW) and surface-acoustic-waves (SAW) devices. Finally, low power consumption and long-lived resonances provided by mechanical resonators make them a suitable choice for on chip oscillators~\cite{campanella_acoustic_2010}. Blending both system ultimately leads to a win-win situation.

Photonic crystals are engineered periodic dielectric structures in which optical waves with wavelengths close to the period of the structure encounter strong dispersion, and in some cases are completely forbidden from propagating within a photonic bandgap frequency window. In the last decade, Si photonics has rapidly developed, in no small part due to the advent of silicon-on-insulator wafer technology in which a thin Si device layer sits atop a low-index insulating oxide layer.  Planar Si photonic crystal circuits may be formed in the top Si device layer, which along with the integrability with micro-electronics, has provided an attractive setting for controlling photons. In a similar way, the periodic patterning of elastic structures can be used to control acoustic wave propagation and create phononic bandgaps.  These structures, known as phononic crystals, have been created in a variety of materials and over a wide range of scales and geometries~\cite{liu_locally_2000,yang_ultrasound_2002,cheng_observation_2006,akimov_hypersonic_2008,gorishnyy_hypersonic_2005,mohammadi_high-q_2009,wen_three-dimensional_2010,gomopoulos_one-dimensional_2010} for applications as diverse as the filtering and focusing of sound\cite{yang_focusing_2004}, the processing of RF/microwave signals~\cite{Ekinci2005,Ekinci2004,Nguyen2007,campanella_acoustic_2010}, and the earthquake proofing of buildings\cite{Gaofeng2010}.  Thin-film OMCs bring together photonic and phononic circuitry~\cite{maldovan_simultaneous_2006,eichenfield_modeling_2009,mohammadi_simultaneous_2010,safavi-naeini_design_2010,Pennec2010}, enabling a new chip-scale platform for delaying, storing, and processing optical and acoustic excitations~\cite{Chang2010,Safavi-Naeini2010a}.  Building on our previous work in quasi-1D OMC systems~\cite{eichenfield_optomechanical_2009}, we demonstrate here a quasi-two-dimensional OMC architecture capable of routing photons and phonons around the full two-dimensional (2D) plane of a Si chip, and enabling complete localization of phonons via a three-dimensional acoustic-wave bandgap.

The phononic crystal used in this work is the recently proposed~\cite{safavi-naeini_design_2010} ``cross'' structure shown in ~Fig.~\ref{Fig1}a. Geometrically, the structure consists of an array of squares connected to each other by thin bridges, or equivalently, a square lattice of cross-shaped holes. The phononic bandgap in this structure arises from the frequency separation between higher frequency tight-binding bands, which have similar frequencies to the resonances of the individual squares, and lower frequency effective-medium bands with frequencies strongly dependent on the width of the connecting bridges, $b = a-h$~\cite{safavi-naeini_design_2010}.  A typical band diagram for a nominal structure ($a=1.265~\mu\text{m}$, $h=1.220~\mu\text{m}$, $w=340~\text{nm}$) is shown in  Fig.~\ref{Fig1}c. Blue (red) lines represent bands with even (odd) vector symmetry for reflections about the $x-y$ plane. The lowest frequency bandgap for the even modes of the simulated cross structure extends from $0.91~\text{GHz}$ to $3.6~\text{GHz}$. Within this bandgap, there are regions of full phononic bandgap (shaded blue) where no mechanical modes of any symmetry exist, and regions of partial symmetry-dependent bandgap (shaded red) where out-of-plane flexural modes with odd symmetry about the $x-y$ plane are allowed. As the bridge width is increased, the lower frequency effective-medium bands become stiffer, causing an increase in their frequency, while the higher frequency tight-binding band frequencies remain essentially constant. A gap-map show in in Fig.~\ref{Fig1}d, showing how the bandgaps in the structure change as a function of phononic crystal bridge width, illustrates this general feature.

As shown in Figs.~\ref{Fig2}a-c, the cross crystal is used as a phononic cage (cavity) for an embedded optical nanocavity~\cite{noda_song_ultra-high-q_2005} (highlighted in a green false color) consisting of a quasi-2D photonic crystal waveguide with a centralized ``defect'' region for localizing photons.  This embedding of an optical cavity within an acoustic cavity enables, through the strong radiation-pressure-coupling of optical and acoustic waves, the probing of the properties of the bandgap-localized phonons via a light field sent through the optical nanocavity.  The theoretical electric field mode profile and the measured high-$Q$ nature of the optical resonance of the photonic crystal cavity are shown in Figs.~\ref{Fig2}d and ~\ref{Fig2}e, respectively.  Such a phonon-photon heterostructure design allows for completely independent tuning of the mechanical and optical properties of our system, and in what follows, we use this feature to probe arrays of structures with different geometric parameters.  In particular, by varying the bridge width $b$ of the outer phononic bandgap crystal, the lower bandgap edge can be swept in frequency and the resulting change in the \emph{lifetime}, \emph{density of states}, and \emph{localization} of the trapped acoustic waves interacting with the central optical cavity can be monitored.  Two different phonon cavity designs, $S_1$ and $S_2$, were fabricated in this study (see Methods). We focus here on the lower acoustic frequency $S_1$ structure, for which an array of devices with bridge width varying from $b=53$~nm to $173$~nm  (in ~$6$~nm increments) was created.  Similar results for the $S_2$ structure are outlined in Appendix~\ref{appB}.

Experimentally we observe the thermally excited acoustic modes of the photonic-phononic crystal through the induced phase-modulation of the optical cavity field~\cite{kippenberg_cavity_2007}.  The mixing of the phase-modulated light from the optical cavity with the transmitted light produces RF/microwave tones upon optical detection with a high speed photodetector (see Methods and Appendix~\ref{appA}). The measured RF-spectra from three different $S_1$ structures with small ($b=57$~nm), medium ($b=106$~nm), and large ($b=160$~nm) bridge widths are shown in Fig.~\ref{Fig3}a. Each narrow tone in the RF-spectra corresponds to a different acoustic resonance interacting with the central optical nanocavity.  Through careful calibration of the optical power and electronic detection, one can extract both the mechanical $Q$-factor (from the linewidth) and the level of optomechanical coupling (from the magnitude of the transduced thermal motion) of each acoustic resonance.  Here we parametrize the strength of the optomechanical coupling by the rate $g$, which corresponds physically to the shift in the optical cavity resonance frequency due to the zero-point motion of the acoustic resonance~\cite{safavi-naeini_design_2010}.  A measurement of the temperature dependence of the acoustic mode spectrum is also performed, and is shown in Fig.~\ref{Fig3}b for one of the acoustic resonances over a temperature range from $300$~K to $10$~K. 

By measuring the entire set of $S_1$ devices in this way, a map may be produced of the localized acoustic modes' properties versus bridge width.  In Figs.~\ref{Fig4}a and ~\ref{Fig4}b we plot the numerically simulated and experimentally measured mode map for the $S_{1}$ structure.  Each marker in these plots corresponds to a different acoustic resonance, with the position of the marker indicating the mode frequency and the size of the marker indicating the mechanical $Q$-factor of the mode (for the numerical simulations all mechanical $Q$-factors above $10^7$ are shown with the same marker size).  Numerical simulations of the optical, mechanical, and optomechanical properties of the structure are performed using the COMSOL~\cite{COMSOL2009} finite-element-method (FEM) software package, with an absorbing boundary condition applied at the exterior of the phononic cage~\cite{eichenfield_modeling_2009,Bindel2005}.  The various bandgap regions are indicated in Fig.~\ref{Fig4}a with the same color coding as in Fig.~\ref{Fig1}c.  Due to the weak radiation pressure coupling to the optical nanocavity of the flexure acoustic modes of odd symmetry about the $x$-$y$ plane of the slab (red mode bands in Fig.~\ref{Fig1}c), we only show in the simulated mode plot of Fig.~\ref{Fig3}a the even symmetry, in-plane acoustic resonances.  

The striking similarity of the simulated and measured mode plots is evidence that the optical nanocavity is able to sensitively probe the in-plane localized acoustic modes of the phononic bandgap structure (the acoustic band with light blue marker in Fig.~\ref{Fig4}a is the one localized in-plane mode which does not show up in the measured plot of Fig.~\ref{Fig4}b; numerical simulations show this mode to be a surface mode at the inner edge of the cross crystal, which does not couple to the central optical cavity).  Within the bandgap, modes are tightly localized (see Fig.~\ref{Fig4}c) and do not radiate acoustic energy, whereas below the bandgap the acoustic modes spread into the exterior cross crystal (see Fig.~\ref{Fig4}d), leaking energy into the surrounding substrate region.  The boundary where the mechanical $Q$-factor drops off is clearly identifiable in the experimentally measured mode plot of Fig.~\ref{Fig4}b (the spectral region below the apparent full phononic bandgap is shaded grey as a guide to the eye), and matches up well with the theoretical lower frequency band-edge of the full phononic bandgap of the cross crystal.  

%Note that this mode is the same as that shown in the theoretical displacement field plots of Figs.~\ref{Fig4}c and \ref{Fig4}d. 

Two other distinguishing features between modes inside and outside a bandgap are the spectral mode density and the strength of the optomechanical coupling.  Below the phononic bandgap, acoustic modes can fill the entire volume of the cross crystal (out to the boundary of the undercut structure where it is finally clamped), resulting in an increase of the mode density (proportional to volume) and a decrease in the optomechanical coupling (proportional to the inverse-square-root of mode volume~\cite{safavi-naeini_design_2010}).  In Fig.~\ref{Fig4}e we have plotted the theoretically computed and experimentally measured values of the optomechancial coupling ($g$) for an acoustic resonance lying near the middle of the full phononic bandgap (this resonance is highlighted in orange in the theoretical plot of Fig.~\ref{Fig4}a and red in the measured plot of Fig.~\ref{Fig4}b).  The measured trend of optomechanical coupling nicely matches that of the theoretical one, and highlights the sharp drop off in optomechanical coupling as the mode crosses the bandgap.  Similarly, in Fig.~\ref{Fig4}f we plot the optomechanical coupling for the acoustic resonances of a single device with bridge width $b \sim 100$~nm (corresonding to a vertical slice in Figs.~\ref{Fig4}a and ~\ref{Fig4}b as indicated by a dashed vertical line).  Again we see good correspondence between theory and experiment, with the drop off in $g$ and the large increase in spectral mode density clearly evident in both plots below the bandgap (note that the frequency position of the bandgap edge is not the same in the theoretical and experimental plots of Figs.~\ref{Fig4}e and ~\ref{Fig4}f due to the slight differences in bridge width).           

Having localized acoustic modes via a full 3D phononic bandgap, and at least in principle removed radiation losses, it is interesting to consider the limits to mechanical damping in these structures.  As shown in Fig.~\ref{Fig3}b for the in-plane acoustic resonance at $1.40~\text{GHz}$ lying well within the phononic bandgap of the cross crystal, the mechanical $Q$-factor increases from a value just above $3000$ at room temperature to a value of $1.3 \times 10^4$ at a temperature of $10$~K.  This temperature dependence of mechanical $Q$ is compared in Appendix~\ref{appC} to Akheiser~\cite{akhieser_original_1939} damping, Landau-Rumer~\cite{landau_rumer_absorption_1937}  damping, a numerical model of thermoelastic damping in the structure~\cite{zener_internal_1938,zener_internal_1938,lifshitz_thermoelastic_2000}, and measurements of acoustic wave attenuation in bulk Si~\cite{Att_Si_lambade_1995}.  The measured temperature dependence and the overall magnitude of the measured mechanical $Q$-factor are seen to be much smaller than any of these comparisons, suggesting that surface effects and/or fabrication-induced damage may be playing an important role in the mechanical damping of the nanosctructed devices studied here.         

Beyond the confinement and localization of acoustic modes in three dimensions, the connected geometry of the 2D-OMC structures presented in this work offers a platform for more complex phonon-photon circuitry.  As has been described in recent theoretical analyses~\cite{safavi-naeini_design_2010,Safavi-Naeini2010a,Chang2010}, such circuitry could be used to create optomechanical systems with greatly enhanced optomechanical coupling, and to realize devices such as traveling wave phonon-photon translators and slow light waveguides~\cite{safavi-naeini_EIT_2010} capable of advanced classical and quantum optical signal processing.  The functionality of these devices are based upon the slow propagation velocity and long relative lifetime of phonons in comparison to photons, which allows for the storage, buffering, and narrowband filtering of optical signals.  In addition, the coupling of optomechanical circuits to a wide variety other physical systems, such as superconducting electronic circuits~\cite{OConnell2010} and atomic vapors~\cite{Kimble_Zoller_Ye_2010}, may also enable the interfacing and networking of different quantum systems.

\section*{Acknowledgements}

This work was supported by the DARPA/MTO ORCHID program through a grant from AFOSR, and the Kavli Nanoscience Institute at Caltech. ASN gratefully acknowledges support from NSERC.

%\newpage
\section*{Methods}
%This is the methods section.  It should contain relevant details that are too detail-oriented for the main text, but should be included in the paper as opposed to in Supplementary Information.  This whole section should be no longer than 300 words.

\noindent\textbf{Fabrication:} Phononic-photonic cavities are fabricated using a Silicon-On-Insulator wafer from SOITEC ($\rho=4$-$20$~$\Omega\cdot$cm, device layer thickness $t=220$~nm, buried-oxide layer thickness $2$~$\mu$m). The cavity geometry is defined by electron beam lithography followed by inductively-coupled-plasma reactive ion etching (ICP-RIE) to transfer the pattern through the $220~\text{nm}$ silicon device layer. The cavities are then undercut using $\text{HF:H}_2\text{O}$ solution to remove the buried oxide layer, and cleaned using a piranha/HF cycle~\cite{borselli_measuringrole_2006}.

\noindent\textbf{Optical measurement technique:} Optically we characterize the samples using a dimpled fiber taper setup connected to a swept-wavelength external-cavity laser~\cite{michael_optical_2007}. A typical fiber taper transmission spectrum is shown in Fig.~\ref{Fig2}(e), with a measured intrinsic optical quality factor of $Q_i=1.5\times10^6$. By controlling the taper position when touching the sample we are able to control the coupling between our fiber taper-probe and the optical cavity. Usually, after touching, the external coupling rate was on the order of tens of MHz ($\kappa_e/2\pi\approx70$~MHz) which corresponds to a transmission dip of $\approx70\%$.

\noindent\textbf{RF measurement technique:}  The transmitted cavity light is sent into an erbium doped fiber amplifier (EDFA) and then onto a high-speed photodetector. The photodetected signal is sent to an oscilloscope ($2$~GHz bandwidth) where the electronic power spectral density (PSD) is computed. Since our devices are in the sideband resolved regime, i.e., the total optical loss rate, $\kappa$, is smaller than the mechanical frequency, $\Omega_\text{M}$, the largest transduced signal is achieved when the laser frequency is detuned from the optical cavity resonance by approximately the mechanical frequency~\cite{eichenfield_optomechanical_2009}. Therefore, for all the measurement, the probe laser was locked $\approx1$~GHz away from the cavity resonance for $S_1$ structures and $\approx2.5$~GHz for $S_2$ structures, both on the blue (higher frequency) side of the cavity.

\bibliographystyle{apsrev}
%\bibliography{../phononic}

\begin{thebibliography}{51}
\expandafter\ifx\csname natexlab\endcsname\relax\def\natexlab#1{#1}\fi
\expandafter\ifx\csname bibnamefont\endcsname\relax
  \def\bibnamefont#1{#1}\fi
\expandafter\ifx\csname bibfnamefont\endcsname\relax
  \def\bibfnamefont#1{#1}\fi
\expandafter\ifx\csname citenamefont\endcsname\relax
  \def\citenamefont#1{#1}\fi
\expandafter\ifx\csname url\endcsname\relax
  \def\url#1{\texttt{#1}}\fi
\expandafter\ifx\csname urlprefix\endcsname\relax\def\urlprefix{URL }\fi
\providecommand{\bibinfo}[2]{#2}
\providecommand{\eprint}[2][]{\url{#2}}

\bibitem[{\citenamefont{Kippenberg et~al.}(2005)\citenamefont{Kippenberg,
  Rokhsari, Carmon, Scherer, and Vahala}}]{Kippenberg2005}
\bibinfo{author}{\bibfnamefont{T.~J.} \bibnamefont{Kippenberg}},
  \bibinfo{author}{\bibfnamefont{H.}~\bibnamefont{Rokhsari}},
  \bibinfo{author}{\bibfnamefont{T.}~\bibnamefont{Carmon}},
  \bibinfo{author}{\bibfnamefont{A.}~\bibnamefont{Scherer}}, \bibnamefont{and}
  \bibinfo{author}{\bibfnamefont{K.~J.} \bibnamefont{Vahala}},
  \bibinfo{journal}{Phys. Rev. Lett.} \textbf{\bibinfo{volume}{95}},
  \bibinfo{pages}{033901–} (\bibinfo{year}{2005}).

\bibitem[{\citenamefont{Carmon et~al.}(2005)\citenamefont{Carmon, Rokhsari,
  Yang, Kippenberg, and Vahala}}]{Carmon2005}
\bibinfo{author}{\bibfnamefont{T.}~\bibnamefont{Carmon}},
  \bibinfo{author}{\bibfnamefont{H.}~\bibnamefont{Rokhsari}},
  \bibinfo{author}{\bibfnamefont{L.}~\bibnamefont{Yang}},
  \bibinfo{author}{\bibfnamefont{T.~J.} \bibnamefont{Kippenberg}},
  \bibnamefont{and} \bibinfo{author}{\bibfnamefont{K.~J.}
  \bibnamefont{Vahala}}, \bibinfo{journal}{Phys. Rev. Lett.}
  \textbf{\bibinfo{volume}{94}}, \bibinfo{pages}{223902}
  (\bibinfo{year}{2005}).

\bibitem[{\citenamefont{Lin et~al.}(2010)\citenamefont{Lin, Rosenberg, Chang,
  Camacho, Eichenfield, Vahala, and Painter}}]{qiang_lin_coherent_2010}
\bibinfo{author}{\bibfnamefont{Q.}~\bibnamefont{Lin}},
  \bibinfo{author}{\bibfnamefont{J.}~\bibnamefont{Rosenberg}},
  \bibinfo{author}{\bibfnamefont{D.}~\bibnamefont{Chang}},
  \bibinfo{author}{\bibfnamefont{R.}~\bibnamefont{Camacho}},
  \bibinfo{author}{\bibfnamefont{M.}~\bibnamefont{Eichenfield}},
  \bibinfo{author}{\bibfnamefont{K.~J.} \bibnamefont{Vahala}},
  \bibnamefont{and} \bibinfo{author}{\bibfnamefont{O.}~\bibnamefont{Painter}},
  \bibinfo{journal}{Nat Photon} \textbf{\bibinfo{volume}{4}},
  \bibinfo{pages}{236} (\bibinfo{year}{2010}).

\bibitem[{\citenamefont{Rosenberg et~al.}(2009)\citenamefont{Rosenberg, Lin,
  and Painter}}]{rosenberg_static_2009}
\bibinfo{author}{\bibfnamefont{J.}~\bibnamefont{Rosenberg}},
  \bibinfo{author}{\bibfnamefont{Q.}~\bibnamefont{Lin}}, \bibnamefont{and}
  \bibinfo{author}{\bibfnamefont{O.}~\bibnamefont{Painter}},
  \bibinfo{journal}{Nat Photon} \textbf{\bibinfo{volume}{3}},
  \bibinfo{pages}{478} (\bibinfo{year}{2009}).

\bibitem[{\citenamefont{Wiederhecker et~al.}(2009)\citenamefont{Wiederhecker,
  Chen, Gondarenko, and Lipson}}]{wiederhecker_controlling_2009}
\bibinfo{author}{\bibfnamefont{G.~S.} \bibnamefont{Wiederhecker}},
  \bibinfo{author}{\bibfnamefont{L.}~\bibnamefont{Chen}},
  \bibinfo{author}{\bibfnamefont{A.}~\bibnamefont{Gondarenko}},
  \bibnamefont{and} \bibinfo{author}{\bibfnamefont{M.}~\bibnamefont{Lipson}},
  \bibinfo{journal}{Nature} \textbf{\bibinfo{volume}{462}},
  \bibinfo{pages}{633} (\bibinfo{year}{2009}).

\bibitem[{\citenamefont{Sridaran and Bhave}(2010)}]{Sridaran2010}
\bibinfo{author}{\bibfnamefont{S.}~\bibnamefont{Sridaran}} \bibnamefont{and}
  \bibinfo{author}{\bibfnamefont{S.~A.} \bibnamefont{Bhave}},
  pp.~\bibinfo{pages}{--} (\bibinfo{year}{2010}).

\bibitem[{\citenamefont{Eichenfield
  et~al.}(2009{\natexlab{a}})\citenamefont{Eichenfield, Camacho, Chan, Vahala,
  and Painter}}]{eichenfield_picogram-_2009}
\bibinfo{author}{\bibfnamefont{M.}~\bibnamefont{Eichenfield}},
  \bibinfo{author}{\bibfnamefont{R.}~\bibnamefont{Camacho}},
  \bibinfo{author}{\bibfnamefont{J.}~\bibnamefont{Chan}},
  \bibinfo{author}{\bibfnamefont{K.~J.} \bibnamefont{Vahala}},
  \bibnamefont{and} \bibinfo{author}{\bibfnamefont{O.}~\bibnamefont{Painter}},
  \bibinfo{journal}{Nature} \textbf{\bibinfo{volume}{459}},
  \bibinfo{pages}{550} (\bibinfo{year}{2009}{\natexlab{a}}).

\bibitem[{\citenamefont{Regal et~al.}(2008)\citenamefont{Regal, Tuefel, and
  Lehnert}}]{Regal08}
\bibinfo{author}{\bibfnamefont{C.~A.} \bibnamefont{Regal}},
  \bibinfo{author}{\bibfnamefont{J.~D.} \bibnamefont{Tuefel}},
  \bibnamefont{and} \bibinfo{author}{\bibfnamefont{K.~W.}
  \bibnamefont{Lehnert}}, \bibinfo{journal}{Nature Physics}
  \textbf{\bibinfo{volume}{4}}, \bibinfo{pages}{555} (\bibinfo{year}{2008}).

\bibitem[{\citenamefont{Dainese et~al.}(2006)\citenamefont{Dainese, Russell,
  Joly, Knight, Wiederhecker, Fragnito, Laude, and
  Khelif}}]{dainese_stimulated_2006}
\bibinfo{author}{\bibfnamefont{P.}~\bibnamefont{Dainese}},
  \bibinfo{author}{\bibfnamefont{P.~S.~J.} \bibnamefont{Russell}},
  \bibinfo{author}{\bibfnamefont{N.}~\bibnamefont{Joly}},
  \bibinfo{author}{\bibfnamefont{J.~C.} \bibnamefont{Knight}},
  \bibinfo{author}{\bibfnamefont{G.~S.} \bibnamefont{Wiederhecker}},
  \bibinfo{author}{\bibfnamefont{H.~L.} \bibnamefont{Fragnito}},
  \bibinfo{author}{\bibfnamefont{V.}~\bibnamefont{Laude}}, \bibnamefont{and}
  \bibinfo{author}{\bibfnamefont{A.}~\bibnamefont{Khelif}},
  \bibinfo{journal}{Nat Phys} \textbf{\bibinfo{volume}{2}},
  \bibinfo{pages}{388} (\bibinfo{year}{2006}).

\bibitem[{\citenamefont{Wiederhecker et~al.}(2008)\citenamefont{Wiederhecker,
  Brenn, Fragnito, and Russell}}]{wiederhecker_coherent_2008}
\bibinfo{author}{\bibfnamefont{G.~S.} \bibnamefont{Wiederhecker}},
  \bibinfo{author}{\bibfnamefont{A.}~\bibnamefont{Brenn}},
  \bibinfo{author}{\bibfnamefont{H.~L.} \bibnamefont{Fragnito}},
  \bibnamefont{and} \bibinfo{author}{\bibfnamefont{P.~S.~J.}
  \bibnamefont{Russell}}, \bibinfo{journal}{Physical Review Letters}
  \textbf{\bibinfo{volume}{100}}, \bibinfo{pages}{203903}
  (\bibinfo{year}{2008}).

\bibitem[{\citenamefont{Eichenfield
  et~al.}(2009{\natexlab{b}})\citenamefont{Eichenfield, Chan, Camacho, Vahala,
  and Painter}}]{eichenfield_optomechanical_2009}
\bibinfo{author}{\bibfnamefont{M.}~\bibnamefont{Eichenfield}},
  \bibinfo{author}{\bibfnamefont{J.}~\bibnamefont{Chan}},
  \bibinfo{author}{\bibfnamefont{R.~M.} \bibnamefont{Camacho}},
  \bibinfo{author}{\bibfnamefont{K.~J.} \bibnamefont{Vahala}},
  \bibnamefont{and} \bibinfo{author}{\bibfnamefont{O.}~\bibnamefont{Painter}},
  \bibinfo{journal}{Nature} \textbf{\bibinfo{volume}{462}}, \bibinfo{pages}{78}
  (\bibinfo{year}{2009}{\natexlab{b}}).

\bibitem[{\citenamefont{Liu et~al.}(2000)\citenamefont{Liu, Zhang, Mao, Zhu,
  Yang, Chan, and Sheng}}]{liu_locally_2000}
\bibinfo{author}{\bibfnamefont{Z.}~\bibnamefont{Liu}},
  \bibinfo{author}{\bibfnamefont{X.}~\bibnamefont{Zhang}},
  \bibinfo{author}{\bibfnamefont{Y.}~\bibnamefont{Mao}},
  \bibinfo{author}{\bibfnamefont{Y.~Y.} \bibnamefont{Zhu}},
  \bibinfo{author}{\bibfnamefont{Z.}~\bibnamefont{Yang}},
  \bibinfo{author}{\bibfnamefont{C.~T.} \bibnamefont{Chan}}, \bibnamefont{and}
  \bibinfo{author}{\bibfnamefont{P.}~\bibnamefont{Sheng}},
  \bibinfo{journal}{Science} \textbf{\bibinfo{volume}{289}},
  \bibinfo{pages}{1734} (\bibinfo{year}{2000}).

\bibitem[{\citenamefont{Yang et~al.}(2002)\citenamefont{Yang, Page, Liu, Cowan,
  Chan, and Sheng}}]{yang_ultrasound_2002}
\bibinfo{author}{\bibfnamefont{S.}~\bibnamefont{Yang}},
  \bibinfo{author}{\bibfnamefont{J.~H.} \bibnamefont{Page}},
  \bibinfo{author}{\bibfnamefont{Z.}~\bibnamefont{Liu}},
  \bibinfo{author}{\bibfnamefont{M.~L.} \bibnamefont{Cowan}},
  \bibinfo{author}{\bibfnamefont{C.~T.} \bibnamefont{Chan}}, \bibnamefont{and}
  \bibinfo{author}{\bibfnamefont{P.}~\bibnamefont{Sheng}},
  \bibinfo{journal}{Physical Review Letters} \textbf{\bibinfo{volume}{88}},
  \bibinfo{pages}{104301} (\bibinfo{year}{2002}).

\bibitem[{\citenamefont{Cheng et~al.}(2006)\citenamefont{Cheng, Wang, Jonas,
  Fytas, and Stefanou}}]{cheng_observation_2006}
\bibinfo{author}{\bibfnamefont{W.}~\bibnamefont{Cheng}},
  \bibinfo{author}{\bibfnamefont{J.}~\bibnamefont{Wang}},
  \bibinfo{author}{\bibfnamefont{U.}~\bibnamefont{Jonas}},
  \bibinfo{author}{\bibfnamefont{G.}~\bibnamefont{Fytas}}, \bibnamefont{and}
  \bibinfo{author}{\bibfnamefont{N.}~\bibnamefont{Stefanou}},
  \bibinfo{journal}{Nat Mater} \textbf{\bibinfo{volume}{5}},
  \bibinfo{pages}{830} (\bibinfo{year}{2006}).

\bibitem[{\citenamefont{Akimov et~al.}(2008)\citenamefont{Akimov, Tanaka,
  Pevtsov, Kaplan, Golubev, Tamura, Yakovlev, and
  Bayer}}]{akimov_hypersonic_2008}
\bibinfo{author}{\bibfnamefont{A.~V.} \bibnamefont{Akimov}},
  \bibinfo{author}{\bibfnamefont{Y.}~\bibnamefont{Tanaka}},
  \bibinfo{author}{\bibfnamefont{A.~B.} \bibnamefont{Pevtsov}},
  \bibinfo{author}{\bibfnamefont{S.~F.} \bibnamefont{Kaplan}},
  \bibinfo{author}{\bibfnamefont{V.~G.} \bibnamefont{Golubev}},
  \bibinfo{author}{\bibfnamefont{S.}~\bibnamefont{Tamura}},
  \bibinfo{author}{\bibfnamefont{D.~R.} \bibnamefont{Yakovlev}},
  \bibnamefont{and} \bibinfo{author}{\bibfnamefont{M.}~\bibnamefont{Bayer}},
  \bibinfo{journal}{Physical Review Letters} \textbf{\bibinfo{volume}{101}},
  \bibinfo{pages}{033902} (\bibinfo{year}{2008}).

\bibitem[{\citenamefont{Gorishnyy et~al.}(2005)\citenamefont{Gorishnyy, Ullal,
  Maldovan, Fytas, and Thomas}}]{gorishnyy_hypersonic_2005}
\bibinfo{author}{\bibfnamefont{T.}~\bibnamefont{Gorishnyy}},
  \bibinfo{author}{\bibfnamefont{C.~K.} \bibnamefont{Ullal}},
  \bibinfo{author}{\bibfnamefont{M.}~\bibnamefont{Maldovan}},
  \bibinfo{author}{\bibfnamefont{G.}~\bibnamefont{Fytas}}, \bibnamefont{and}
  \bibinfo{author}{\bibfnamefont{E.~L.} \bibnamefont{Thomas}},
  \bibinfo{journal}{Physical Review Letters} \textbf{\bibinfo{volume}{94}},
  \bibinfo{pages}{115501} (\bibinfo{year}{2005}).

\bibitem[{\citenamefont{Mohammadi et~al.}(2009)\citenamefont{Mohammadi,
  Eftekhar, Hunt, and Adibi}}]{mohammadi_high-q_2009}
\bibinfo{author}{\bibfnamefont{S.}~\bibnamefont{Mohammadi}},
  \bibinfo{author}{\bibfnamefont{A.~A.} \bibnamefont{Eftekhar}},
  \bibinfo{author}{\bibfnamefont{W.~D.} \bibnamefont{Hunt}}, \bibnamefont{and}
  \bibinfo{author}{\bibfnamefont{A.}~\bibnamefont{Adibi}},
  \bibinfo{journal}{Applied Physics Letters} \textbf{\bibinfo{volume}{94}},
  \bibinfo{pages}{051906} (\bibinfo{year}{2009}).

\bibitem[{\citenamefont{Wen et~al.}(2010)\citenamefont{Wen, Sun, Dais,
  Grützmacher, Wu, Shi, and Sun}}]{wen_three-dimensional_2010}
\bibinfo{author}{\bibfnamefont{Y.}~\bibnamefont{Wen}},
  \bibinfo{author}{\bibfnamefont{J.}~\bibnamefont{Sun}},
  \bibinfo{author}{\bibfnamefont{C.}~\bibnamefont{Dais}},
  \bibinfo{author}{\bibfnamefont{D.}~\bibnamefont{Grützmacher}},
  \bibinfo{author}{\bibfnamefont{T.}~\bibnamefont{Wu}},
  \bibinfo{author}{\bibfnamefont{J.}~\bibnamefont{Shi}}, \bibnamefont{and}
  \bibinfo{author}{\bibfnamefont{C.}~\bibnamefont{Sun}},
  \bibinfo{journal}{Applied Physics Letters} \textbf{\bibinfo{volume}{96}},
  \bibinfo{pages}{123113} (\bibinfo{year}{2010}).

\bibitem[{\citenamefont{Gomopoulos et~al.}(2010)\citenamefont{Gomopoulos,
  Maschke, Koh, Thomas, Tremel, Butt, and
  Fytas}}]{gomopoulos_one-dimensional_2010}
\bibinfo{author}{\bibfnamefont{N.}~\bibnamefont{Gomopoulos}},
  \bibinfo{author}{\bibfnamefont{D.}~\bibnamefont{Maschke}},
  \bibinfo{author}{\bibfnamefont{C.~Y.} \bibnamefont{Koh}},
  \bibinfo{author}{\bibfnamefont{E.~L.} \bibnamefont{Thomas}},
  \bibinfo{author}{\bibfnamefont{W.}~\bibnamefont{Tremel}},
  \bibinfo{author}{\bibfnamefont{H.}~\bibnamefont{Butt}}, \bibnamefont{and}
  \bibinfo{author}{\bibfnamefont{G.}~\bibnamefont{Fytas}},
  \bibinfo{journal}{Nano Letters} \textbf{\bibinfo{volume}{10}},
  \bibinfo{pages}{980} (\bibinfo{year}{2010}).

\bibitem[{\citenamefont{Yang et~al.}(2004)\citenamefont{Yang, Page, Liu, Cowan,
  Chan, and Sheng}}]{yang_focusing_2004}
\bibinfo{author}{\bibfnamefont{S.}~\bibnamefont{Yang}},
  \bibinfo{author}{\bibfnamefont{J.~H.} \bibnamefont{Page}},
  \bibinfo{author}{\bibfnamefont{Z.}~\bibnamefont{Liu}},
  \bibinfo{author}{\bibfnamefont{M.~L.} \bibnamefont{Cowan}},
  \bibinfo{author}{\bibfnamefont{C.~T.} \bibnamefont{Chan}}, \bibnamefont{and}
  \bibinfo{author}{\bibfnamefont{P.}~\bibnamefont{Sheng}},
  \bibinfo{journal}{Physical Review Letters} \textbf{\bibinfo{volume}{93}},
  \bibinfo{pages}{024301} (\bibinfo{year}{2004}).

\bibitem[{\citenamefont{Ekinci and Roukes}(2005)}]{Ekinci2005}
\bibinfo{author}{\bibfnamefont{K.~L.} \bibnamefont{Ekinci}} \bibnamefont{and}
  \bibinfo{author}{\bibfnamefont{M.~L.} \bibnamefont{Roukes}},
  \bibinfo{journal}{Rev. Sci. Instrum.} \textbf{\bibinfo{volume}{76}},
  \bibinfo{pages}{061101} (\bibinfo{year}{2005}), ISSN
  \bibinfo{issn}{00346748}.

\bibitem[{\citenamefont{Ekinci et~al.}(2004)\citenamefont{Ekinci, Yang, and
  Roukes}}]{Ekinci2004}
\bibinfo{author}{\bibfnamefont{K.~L.} \bibnamefont{Ekinci}},
  \bibinfo{author}{\bibfnamefont{Y.~T.} \bibnamefont{Yang}}, \bibnamefont{and}
  \bibinfo{author}{\bibfnamefont{M.~L.} \bibnamefont{Roukes}},
  \bibinfo{journal}{J. Appl. Phys.} \textbf{\bibinfo{volume}{95}},
  \bibinfo{pages}{2682} (\bibinfo{year}{2004}), ISSN \bibinfo{issn}{00218979}.

\bibitem[{\citenamefont{Nguyen}(2007)}]{Nguyen2007}
\bibinfo{author}{\bibfnamefont{C.~T.~C.} \bibnamefont{Nguyen}},
  \bibinfo{journal}{IEEE Trans Ultrason Ferroelectr Freq Control}
  \textbf{\bibinfo{volume}{54}}, \bibinfo{pages}{251} (\bibinfo{year}{2007}),
  ISSN \bibinfo{issn}{0885-3010}.

\bibitem[{\citenamefont{Campanella}(2010)}]{campanella_acoustic_2010}
\bibinfo{author}{\bibfnamefont{H.}~\bibnamefont{Campanella}},
  \emph{\bibinfo{title}{Acoustic Wave and Electromechanical Resonators: Concept
  to Key Applications}} (\bibinfo{publisher}{Artech House Publishers},
  \bibinfo{year}{2010}), ISBN \bibinfo{isbn}{1607839776}.

\bibitem[{\citenamefont{Gaofeng and Zhifei}(2010)}]{Gaofeng2010}
\bibinfo{author}{\bibfnamefont{J.}~\bibnamefont{Gaofeng}} \bibnamefont{and}
  \bibinfo{author}{\bibfnamefont{S.}~\bibnamefont{Zhifei}},
  \bibinfo{journal}{{Earthq. Eng. \& Eng. Vib.}} \textbf{\bibinfo{volume}{9}},
  \bibinfo{pages}{75} (\bibinfo{year}{2010}).

\bibitem[{\citenamefont{Maldovan and
  Thomas}(2006)}]{maldovan_simultaneous_2006}
\bibinfo{author}{\bibfnamefont{M.}~\bibnamefont{Maldovan}} \bibnamefont{and}
  \bibinfo{author}{\bibfnamefont{E.~L.} \bibnamefont{Thomas}},
  \bibinfo{journal}{Applied Physics Letters} \textbf{\bibinfo{volume}{88}},
  \bibinfo{pages}{251907} (\bibinfo{year}{2006}).

\bibitem[{\citenamefont{Eichenfield
  et~al.}(2009{\natexlab{c}})\citenamefont{Eichenfield, Chan, {Safavi-Naeini},
  Vahala, and Painter}}]{eichenfield_modeling_2009}
\bibinfo{author}{\bibfnamefont{M.}~\bibnamefont{Eichenfield}},
  \bibinfo{author}{\bibfnamefont{J.}~\bibnamefont{Chan}},
  \bibinfo{author}{\bibfnamefont{A.~H.} \bibnamefont{{Safavi-Naeini}}},
  \bibinfo{author}{\bibfnamefont{K.~J.} \bibnamefont{Vahala}},
  \bibnamefont{and} \bibinfo{author}{\bibfnamefont{O.}~\bibnamefont{Painter}},
  \bibinfo{journal}{Optics Express} \textbf{\bibinfo{volume}{17}},
  \bibinfo{pages}{20078} (\bibinfo{year}{2009}{\natexlab{c}}).

\bibitem[{\citenamefont{Mohammadi et~al.}(2010)\citenamefont{Mohammadi,
  Eftekhar, Khelif, and Adibi}}]{mohammadi_simultaneous_2010}
\bibinfo{author}{\bibfnamefont{S.}~\bibnamefont{Mohammadi}},
  \bibinfo{author}{\bibfnamefont{A.~A.} \bibnamefont{Eftekhar}},
  \bibinfo{author}{\bibfnamefont{A.}~\bibnamefont{Khelif}}, \bibnamefont{and}
  \bibinfo{author}{\bibfnamefont{A.}~\bibnamefont{Adibi}},
  \bibinfo{journal}{Optics Express} \textbf{\bibinfo{volume}{18}},
  \bibinfo{pages}{9164} (\bibinfo{year}{2010}).

\bibitem[{\citenamefont{{Safavi-Naeini} and
  Painter}(2010)}]{safavi-naeini_design_2010}
\bibinfo{author}{\bibfnamefont{A.~H.} \bibnamefont{{Safavi-Naeini}}}
  \bibnamefont{and} \bibinfo{author}{\bibfnamefont{O.}~\bibnamefont{Painter}},
  \bibinfo{journal}{Optics Express} \textbf{\bibinfo{volume}{18}},
  \bibinfo{pages}{14926} (\bibinfo{year}{2010}).

\bibitem[{\citenamefont{Pennec et~al.}(2010)\citenamefont{Pennec, Rouhani,
  El~Boudouti, Li, El~Hassouani, Vasseur, Papanikolaou, Benchabane, Laude, and
  Martinez}}]{Pennec2010}
\bibinfo{author}{\bibfnamefont{Y.}~\bibnamefont{Pennec}},
  \bibinfo{author}{\bibfnamefont{B.~D.} \bibnamefont{Rouhani}},
  \bibinfo{author}{\bibfnamefont{E.~H.} \bibnamefont{El~Boudouti}},
  \bibinfo{author}{\bibfnamefont{C.}~\bibnamefont{Li}},
  \bibinfo{author}{\bibfnamefont{Y.}~\bibnamefont{El~Hassouani}},
  \bibinfo{author}{\bibfnamefont{J.~O.} \bibnamefont{Vasseur}},
  \bibinfo{author}{\bibfnamefont{N.}~\bibnamefont{Papanikolaou}},
  \bibinfo{author}{\bibfnamefont{S.}~\bibnamefont{Benchabane}},
  \bibinfo{author}{\bibfnamefont{V.}~\bibnamefont{Laude}}, \bibnamefont{and}
  \bibinfo{author}{\bibfnamefont{A.}~\bibnamefont{Martinez}},
  \bibinfo{journal}{Opt. Express} \textbf{\bibinfo{volume}{18}},
  \bibinfo{pages}{14301} (\bibinfo{year}{2010}).

\bibitem[{\citenamefont{Chang et~al.}(2010)\citenamefont{Chang,
  {Safavi-Naeini}, Hafezi, and Painter}}]{Chang2010}
\bibinfo{author}{\bibfnamefont{D.}~\bibnamefont{Chang}},
  \bibinfo{author}{\bibfnamefont{A.~H.} \bibnamefont{{Safavi-Naeini}}},
  \bibinfo{author}{\bibfnamefont{M.}~\bibnamefont{Hafezi}}, \bibnamefont{and}
  \bibinfo{author}{\bibfnamefont{O.}~\bibnamefont{Painter}},
  \bibinfo{journal}{arXiv:1006.3829}  (\bibinfo{year}{2010}).

\bibitem[{\citenamefont{Safavi-Naeini and Painter}(2010)}]{Safavi-Naeini2010a}
\bibinfo{author}{\bibfnamefont{A.~H.} \bibnamefont{Safavi-Naeini}}
  \bibnamefont{and} \bibinfo{author}{\bibfnamefont{O.}~\bibnamefont{Painter}}
  (\bibinfo{year}{2010}), \eprint{arXiv:1009.3529}.

\bibitem[{\citenamefont{Song et~al.}(2005)\citenamefont{Song, Noda, Asano, and
  Akahane}}]{noda_song_ultra-high-q_2005}
\bibinfo{author}{\bibfnamefont{B.}~\bibnamefont{Song}},
  \bibinfo{author}{\bibfnamefont{S.}~\bibnamefont{Noda}},
  \bibinfo{author}{\bibfnamefont{T.}~\bibnamefont{Asano}}, \bibnamefont{and}
  \bibinfo{author}{\bibfnamefont{Y.}~\bibnamefont{Akahane}},
  \bibinfo{journal}{Nat Mater} \textbf{\bibinfo{volume}{4}},
  \bibinfo{pages}{207} (\bibinfo{year}{2005}).

\bibitem[{\citenamefont{Kippenberg and Vahala}(2007)}]{kippenberg_cavity_2007}
\bibinfo{author}{\bibfnamefont{T.~J.} \bibnamefont{Kippenberg}}
  \bibnamefont{and} \bibinfo{author}{\bibfnamefont{K.~J.}
  \bibnamefont{Vahala}}, \bibinfo{journal}{Optics Express}
  \textbf{\bibinfo{volume}{15}}, \bibinfo{pages}{17172} (\bibinfo{year}{2007}).

\bibitem[{COM(2009)}]{COMSOL2009}
\emph{\bibinfo{title}{COMSOL Multphysics 3.5}} (\bibinfo{year}{2009}).

\bibitem[{\citenamefont{Bindel and Govindjee}(2005)}]{Bindel2005}
\bibinfo{author}{\bibfnamefont{D.~S.} \bibnamefont{Bindel}} \bibnamefont{and}
  \bibinfo{author}{\bibfnamefont{S.}~\bibnamefont{Govindjee}},
  \bibinfo{journal}{International Journal for Numerical Methods in Engineering}
  \textbf{\bibinfo{volume}{64}}, \bibinfo{pages}{789} (\bibinfo{year}{2005}).

\bibitem[{\citenamefont{Akhieser}(1939)}]{akhieser_original_1939}
\bibinfo{author}{\bibfnamefont{A.}~\bibnamefont{Akhieser}},
  \bibinfo{journal}{J. Phys. {(Moscow)}} \textbf{\bibinfo{volume}{1}},
  \bibinfo{pages}{277} (\bibinfo{year}{1939}).

\bibitem[{\citenamefont{Landau and Rumer}(1937)}]{landau_rumer_absorption_1937}
\bibinfo{author}{\bibfnamefont{L.}~\bibnamefont{Landau}} \bibnamefont{and}
  \bibinfo{author}{\bibfnamefont{G.}~\bibnamefont{Rumer}},
  \bibinfo{journal}{Phys. Z. Sowjetunion} \textbf{\bibinfo{volume}{11}},
  \bibinfo{pages}{18} (\bibinfo{year}{1937}).

\bibitem[{\citenamefont{Zener}(1938)}]{zener_internal_1938}
\bibinfo{author}{\bibfnamefont{C.}~\bibnamefont{Zener}},
  \bibinfo{journal}{Physical Review} \textbf{\bibinfo{volume}{53}},
  \bibinfo{pages}{90} (\bibinfo{year}{1938}).

\bibitem[{\citenamefont{Lifshitz and
  Roukes}(2000)}]{lifshitz_thermoelastic_2000}
\bibinfo{author}{\bibfnamefont{R.}~\bibnamefont{Lifshitz}} \bibnamefont{and}
  \bibinfo{author}{\bibfnamefont{M.~L.} \bibnamefont{Roukes}},
  \bibinfo{journal}{Physical Review B} \textbf{\bibinfo{volume}{61}},
  \bibinfo{pages}{5600} (\bibinfo{year}{2000}).

\bibitem[{\citenamefont{Lambade et~al.}(1995)\citenamefont{Lambade,
  Sahasrabudhe, and Rajagopalan}}]{Att_Si_lambade_1995}
\bibinfo{author}{\bibfnamefont{S.~D.} \bibnamefont{Lambade}},
  \bibinfo{author}{\bibfnamefont{G.~G.} \bibnamefont{Sahasrabudhe}},
  \bibnamefont{and}
  \bibinfo{author}{\bibfnamefont{S.}~\bibnamefont{Rajagopalan}},
  \bibinfo{journal}{Physical Review B} \textbf{\bibinfo{volume}{51}},
  \bibinfo{pages}{15861} (\bibinfo{year}{1995}),
  \urlprefix\url{http://link.aps.org/doi/10.1103/PhysRevB.51.15861}.

\bibitem[{\citenamefont{Safavi-Naeini et~al.}(2010)\citenamefont{Safavi-Naeini,
  Alegre, Chan, Eichenfield, Winger, Lin, Hill, Chang, and
  Painter}}]{safavi-naeini_EIT_2010}
\bibinfo{author}{\bibfnamefont{A.~H.} \bibnamefont{Safavi-Naeini}},
  \bibinfo{author}{\bibfnamefont{T.~P.~M.} \bibnamefont{Alegre}},
  \bibinfo{author}{\bibfnamefont{J.}~\bibnamefont{Chan}},
  \bibinfo{author}{\bibfnamefont{M.}~\bibnamefont{Eichenfield}},
  \bibinfo{author}{\bibfnamefont{M.}~\bibnamefont{Winger}},
  \bibinfo{author}{\bibfnamefont{Q.}~\bibnamefont{Lin}},
  \bibinfo{author}{\bibfnamefont{J.~T.} \bibnamefont{Hill}},
  \bibinfo{author}{\bibfnamefont{D.}~\bibnamefont{Chang}}, \bibnamefont{and}
  \bibinfo{author}{\bibfnamefont{O.}~\bibnamefont{Painter}}
  (\bibinfo{year}{2010}), \eprint{arXiv:1012.1934}.

\bibitem[{\citenamefont{O'Connell et~al.}(2010)\citenamefont{O'Connell,
  Hofheinz, Ansmann, Bialczak, Lenander, Lucero, Neeley, Sank, Wang, Weides
  et~al.}}]{OConnell2010}
\bibinfo{author}{\bibfnamefont{A.~D.} \bibnamefont{O'Connell}},
  \bibinfo{author}{\bibfnamefont{M.}~\bibnamefont{Hofheinz}},
  \bibinfo{author}{\bibfnamefont{M.}~\bibnamefont{Ansmann}},
  \bibinfo{author}{\bibfnamefont{R.~C.} \bibnamefont{Bialczak}},
  \bibinfo{author}{\bibfnamefont{M.}~\bibnamefont{Lenander}},
  \bibinfo{author}{\bibfnamefont{E.}~\bibnamefont{Lucero}},
  \bibinfo{author}{\bibfnamefont{M.}~\bibnamefont{Neeley}},
  \bibinfo{author}{\bibfnamefont{D.}~\bibnamefont{Sank}},
  \bibinfo{author}{\bibfnamefont{H.}~\bibnamefont{Wang}},
  \bibinfo{author}{\bibfnamefont{M.}~\bibnamefont{Weides}},
  \bibnamefont{et~al.}, \bibinfo{journal}{Nature}
  \textbf{\bibinfo{volume}{464}}, \bibinfo{pages}{697} (\bibinfo{year}{2010}),
  ISSN \bibinfo{issn}{0028-0836},
  \urlprefix\url{http://dx.doi.org/10.1038/nature08967}.

\bibitem[{\citenamefont{Wallquist et~al.}(2010)\citenamefont{Wallquist,
  Hammerer, Zoller, Genes, Ludwig, Marquardt, Treutlein, Ye, and
  Kimble}}]{Kimble_Zoller_Ye_2010}
\bibinfo{author}{\bibfnamefont{M.}~\bibnamefont{Wallquist}},
  \bibinfo{author}{\bibfnamefont{K.}~\bibnamefont{Hammerer}},
  \bibinfo{author}{\bibfnamefont{P.}~\bibnamefont{Zoller}},
  \bibinfo{author}{\bibfnamefont{C.}~\bibnamefont{Genes}},
  \bibinfo{author}{\bibfnamefont{M.}~\bibnamefont{Ludwig}},
  \bibinfo{author}{\bibfnamefont{F.}~\bibnamefont{Marquardt}},
  \bibinfo{author}{\bibfnamefont{P.}~\bibnamefont{Treutlein}},
  \bibinfo{author}{\bibfnamefont{J.}~\bibnamefont{Ye}}, \bibnamefont{and}
  \bibinfo{author}{\bibfnamefont{H.~J.} \bibnamefont{Kimble}},
  \bibinfo{journal}{Phys. Rev. A} \textbf{\bibinfo{volume}{81}},
  \bibinfo{pages}{023816} (\bibinfo{year}{2010}).

\bibitem[{\citenamefont{Borselli et~al.}(2006)\citenamefont{Borselli, Johnson,
  and Painter}}]{borselli_measuringrole_2006}
\bibinfo{author}{\bibfnamefont{M.}~\bibnamefont{Borselli}},
  \bibinfo{author}{\bibfnamefont{T.~J.} \bibnamefont{Johnson}},
  \bibnamefont{and} \bibinfo{author}{\bibfnamefont{O.}~\bibnamefont{Painter}},
  \bibinfo{journal}{Applied Physics Letters} \textbf{\bibinfo{volume}{88}},
  \bibinfo{pages}{131114} (\bibinfo{year}{2006}).

\bibitem[{\citenamefont{Michael et~al.}(2007)\citenamefont{Michael, Borselli,
  Johnson, Chrystal, and Painter}}]{michael_optical_2007}
\bibinfo{author}{\bibfnamefont{C.~P.} \bibnamefont{Michael}},
  \bibinfo{author}{\bibfnamefont{M.}~\bibnamefont{Borselli}},
  \bibinfo{author}{\bibfnamefont{T.~J.} \bibnamefont{Johnson}},
  \bibinfo{author}{\bibfnamefont{C.}~\bibnamefont{Chrystal}}, \bibnamefont{and}
  \bibinfo{author}{\bibfnamefont{O.}~\bibnamefont{Painter}},
  \bibinfo{journal}{Optics Express} \textbf{\bibinfo{volume}{15}},
  \bibinfo{pages}{4745} (\bibinfo{year}{2007}).

\bibitem[{\citenamefont{Zener}(1937)}]{zener_internal_1937}
\bibinfo{author}{\bibfnamefont{C.}~\bibnamefont{Zener}},
  \bibinfo{journal}{Physical Review} \textbf{\bibinfo{volume}{52}},
  \bibinfo{pages}{230} (\bibinfo{year}{1937}).

\bibitem[{\citenamefont{Woodruff and
  Ehrenreich}(1961)}]{woodruff_absorption_sound_1961}
\bibinfo{author}{\bibfnamefont{T.~O.} \bibnamefont{Woodruff}} \bibnamefont{and}
  \bibinfo{author}{\bibfnamefont{H.}~\bibnamefont{Ehrenreich}},
  \bibinfo{journal}{Physical Review} \textbf{\bibinfo{volume}{123}},
  \bibinfo{pages}{1553} (\bibinfo{year}{1961}),
  \urlprefix\url{http://link.aps.org/doi/10.1103/PhysRev.123.1553}.

\bibitem[{\citenamefont{Philip and
  Breazeale}(1983)}]{Gruneiser_philip_third-order_1983}
\bibinfo{author}{\bibfnamefont{J.}~\bibnamefont{Philip}} \bibnamefont{and}
  \bibinfo{author}{\bibfnamefont{M.~A.} \bibnamefont{Breazeale}},
  \bibinfo{journal}{Journal of Applied Physics} \textbf{\bibinfo{volume}{54}},
  \bibinfo{pages}{752} (\bibinfo{year}{1983}), ISSN \bibinfo{issn}{00218979},
  \urlprefix\url{http://link.aip.org/link/JAPIAU/v54/i2/p752/s1&Agg=doi}.

\bibitem[{\citenamefont{Estreicher et~al.}(2004)\citenamefont{Estreicher,
  Sanati, West, and Ruymgaart}}]{Cv_estreicher_thermodynamics_2004}
\bibinfo{author}{\bibfnamefont{S.~K.} \bibnamefont{Estreicher}},
  \bibinfo{author}{\bibfnamefont{M.}~\bibnamefont{Sanati}},
  \bibinfo{author}{\bibfnamefont{D.}~\bibnamefont{West}}, \bibnamefont{and}
  \bibinfo{author}{\bibfnamefont{F.}~\bibnamefont{Ruymgaart}},
  \bibinfo{journal}{Physical Review B} \textbf{\bibinfo{volume}{70}},
  \bibinfo{pages}{125209} (\bibinfo{year}{2004}),
  \urlprefix\url{http://link.aps.org/doi/10.1103/PhysRevB.70.125209}.

\bibitem[{\citenamefont{Duwel et~al.}(2006)\citenamefont{Duwel, Candler, Kenny,
  and Varghese}}]{duwel_engineering_2006}
\bibinfo{author}{\bibfnamefont{A.}~\bibnamefont{Duwel}},
  \bibinfo{author}{\bibfnamefont{R.}~\bibnamefont{Candler}},
  \bibinfo{author}{\bibfnamefont{T.}~\bibnamefont{Kenny}}, \bibnamefont{and}
  \bibinfo{author}{\bibfnamefont{M.}~\bibnamefont{Varghese}},
  \bibinfo{journal}{Journal of Microelectromechanical Systems}
  \textbf{\bibinfo{volume}{15}}, \bibinfo{pages}{1437} (\bibinfo{year}{2006}).

\end{thebibliography}

\appendix

\section{Experimental Setup}
\label{appA}
\begin{figure*}[ht!]
\includegraphics[width=2\columnwidth]{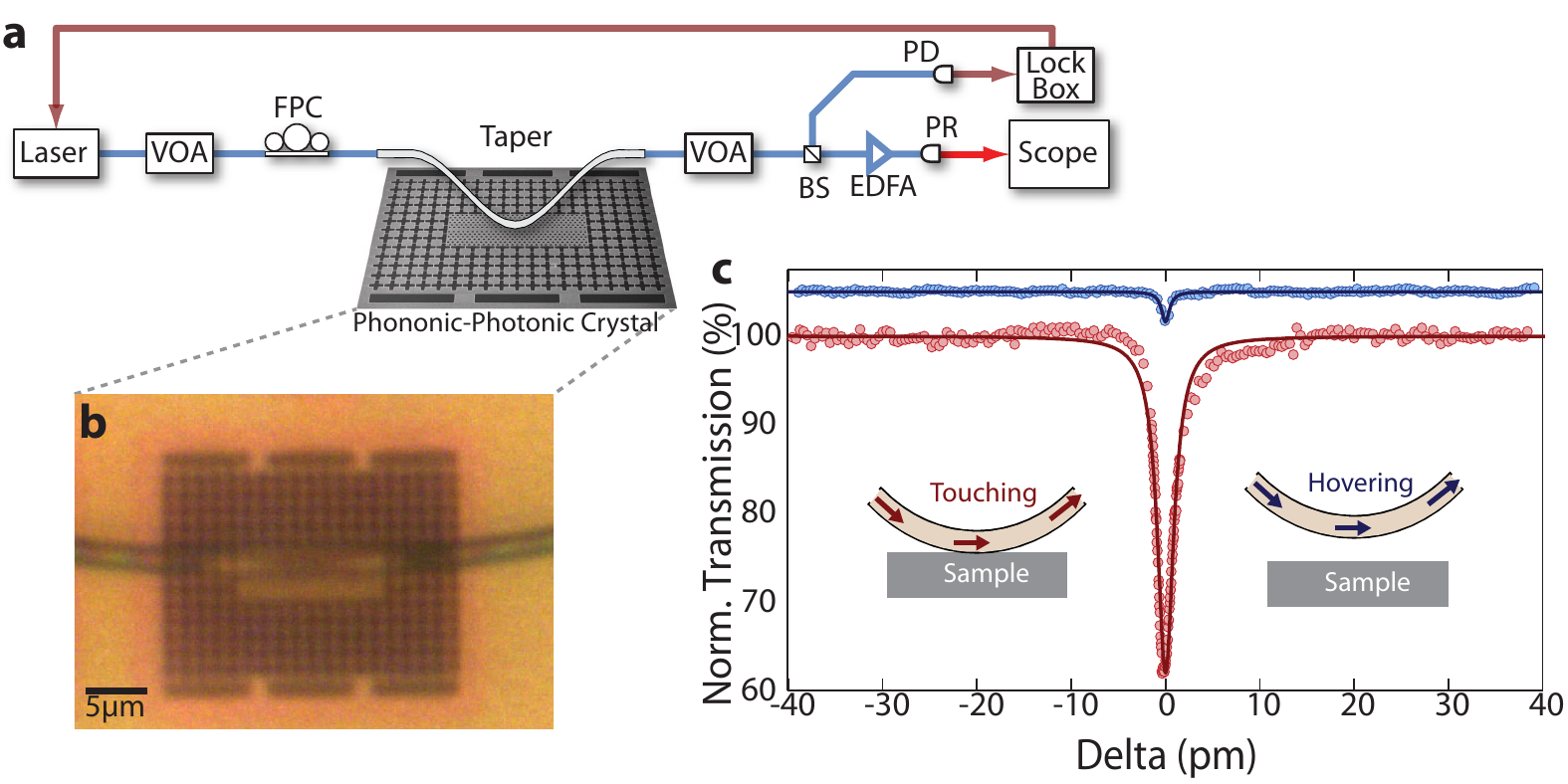} %single column figure - for two-column format
\caption{\label{Fig_ExpSetup} \textbf{Experimental Setup} \textbf{a}, Experimental setup for measuring the PSD. \textbf{b}, Optical micrograph of the tapered fiber coupled to the device while performing experiments. \textbf{c}, Optical spectra for two different positions of the taper relative to the device are shown.}
\end{figure*}

The experimental setup used to measure the phononic-photonic crystal cavity properties is shown in Fig.~\ref{Fig_ExpSetup}(a). A fiber-coupled tunable infrared laser, (New Focus Velocity, model TLB-6328) spanning approximately $60$~nm, centered around $1540$~nm, has its intensity and polarization controlled respectively by a variable optical attenuator (VOA) and a fiber polarization controller (FPC). The laser light is coupled to a tapered, dimpled optical fiber (Taper) which has its position controlled with nanometer-scale precision. The transmission from the fiber is passed through another VOA before being detected.

To measure the optical properties, a photodetector (PD, New Focus Nanosecond Photodetector, model 1623) is used. The detected optical transmission signal is recorded while sweeping the laser frequency. By controlling the distance between the fiber taper and the sample, the external coupling rate ($\kappa_e$) is changed. Fig.~\ref{Fig_ExpSetup}(c) shows the change in the coupling rate for two different positions of the fiber taper. In the limit where the external coupling rate is zero we can measure the intrinsic coupling rate ($\kappa_i$). The total optical loss is then $\kappa=\kappa_e+\kappa_i$.

%%%%%%%%%%%%%%%%%%%%%%%%%%%%%%%%%%% FIGURE 01 %%%%%%%%%%%%%%%%%%%%%%%%%%%%%%%%%%%%%%%%%%%%%%%%%%%%%%%%
%%%%%%%%%%%%%%%%%%%%%%%%%%%%%%%%%%%%%%% FIGURE 01 %%%%%%%%%%%%%%%%%%%%%%%%%%%%%%%%%%%%%%%%%%%%%%%%%%%%%%%%

To measure the mechanical properties, the transmitted signal is sent through an erbium doped fiber amplifier (EDFA) and  sent to a high-speed photoreceiver (PR, New Focus model, 1554-B) with a maximum transimpedance gain of $1,000$~V/A and a bandwidth ($3$~dB rolloff point) of $12$~GHz. The RF voltage from the photoreceiver is  connected to the $50~\Omega$ input impedance of the oscilloscope. The oscilloscope can perform a Fast Fourier Transform (FFT) to yield the RF power spectral density (RF PSD). The RF PSD is calibrated using a frequency generator that outputs a variable frequency sinusoid with known power.

As stated in the main text, our devices are in the sideband resolved limit, i.e. the total optical loss rate is smaller than the mechanical frequency, $\kappa < \Omega_\text{M}$. Therefore the largest transduced signal is achieved when the laser frequency detuned from the optical cavity resonance by the mechanical frequency~\cite{eichenfield_optomechanical_2009}. The probe laser is locked to approximately $1~\text{GHz}$ ($2.5~\text{GHz}$) on the blue side of the cavity resonance for the $S_1$ ($S_2$) structures. By measuring the transmission contrast during the acquisition of the RF PSD and comparing with the transmission curve of each device (as shown in Fig.~\ref{Fig_ExpSetup}(c)) we determine the laser detuning and the dropped power into the cavity.  To lock the probe laser frequency a given frequency away from the resonance, a $90/10$ beam splitter (BS) is added to the optical path, and the signal from the $10\%$ arm is feed to a PD connected to a locking circuit which compares the voltage level from the transmission signal to a predetermined value to generate an error signal. The error signal is then fed into the laser to stabilize the laser frequency.

\section{Phononic Band Gap Tuning via Lattice Size}
\label{appB}
%%%%%%%%%%%%%%%%%%%%%%%%%%%%%%%%%%% FIGURE 02 %%%%%%%%%%%%%%%%%%%%%%%%%%%%%%%%%%%%%%%%%%%%%%%%%%%%%%%%
\begin{figure*}[ht!]
\includegraphics[width=2\columnwidth]{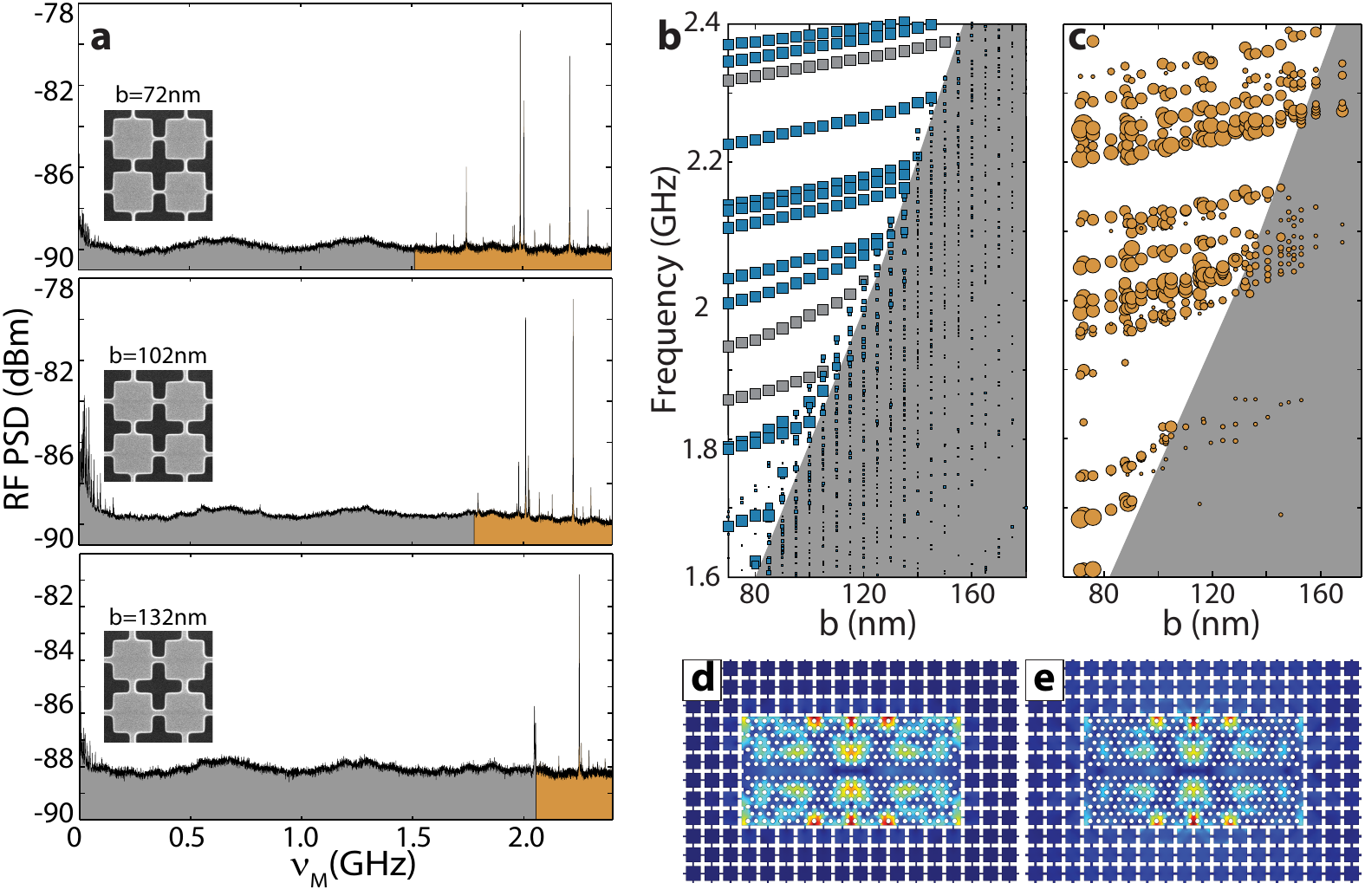} %single column figure - for two-column format
\caption{\label{Fig_Nt4} \textbf{$S_2$ cavity design}. \textbf{a}, RF-power spectral density for three devices with different bridge widths, $b$. The gray shaded area refers to regions outside of the full bandgap while the orange shaded area are for frequencies within the bandgap.  \textbf{b}, results of full 3D-FEM simulations of phononic localized modes as a function of bridge width $b$. For each bridge-width a full 3D-FEM simulation with an absorbing perfect-matched-layer (PML) is performed. Each square corresponds to a single mode, with its size is proportional to the logarithm of the mechanical $Q$-factor. The gray shaded squares represent modes which are optically dark ($g$'s small enough to be lower than our detection noise floor).  \textbf{c}, Measurement ($\circ$) results of the localized mechanical modes for the a set of fabricated $S_2$ samples. Each bridge size $b$ was measured by careful analysis of scanning electron micrographs. The marker sizes are proportional to measured mechanical $Q$-factor. \textbf{d} and \textbf{e}, FEM simulations of the displacement field amplitude $|\mathbf{Q}(\mathbf{r})|$ for the mechanical mode at $2.1$GHz shown in \textbf{b}. In \textbf{d}, the mode is within the phononic bandgap resulting in a radiation limited $Q_\text{M}^\text{(rad)}\approx10^9$. When on the edge of the bandgap, the mode is less localized as shown in \textbf{e}, and has a reduced $Q_\text{M}^\text{(rad)} < 10^3$.}
\end{figure*}
%%%%%%%%%%%%%%%%%%%%%%%%%%%%%%%%%%%%%%% FIGURE 02 %%%%%%%%%%%%%%%%%%%%%%%%%%%%%%%%%%%%%%%%%%%%%%%%%%%%

In the main text we demonstrate a phononic bandgap around $1.3$~GHz based upon the $S_1$ design.  In this section we present measurements performed on a second design ($S_{2}$) with phononic bandgap in a different frequency range.  Following the nomenclature in Fig.~1 of the main text, the nominal dimensions for the $S_2$ devices are: $a=925$~nm, $h=850$~nm, and $w=210$~nm, which allows for an acoustic bandgap around $1.6$ to $2.5$~GHz.  The optical nanocavity in the $S_2$ devices are the same as that used in $S_{1}$ devices.

To show the presence of a bandgap, we fabricate and measure a set of devices with bridge widths varying from $70$ to $170$~nm. The RF PSD of the optically transmitted signal is shown in Fig.~\ref{Fig_Nt4}a for three different $S_{2}$ structures with different bridge widths.  By comparing the spectra for different bridge width the presence of a phononic bandgap becomes evident.  Figs.~\ref{Fig_Nt4}b and \ref{Fig_Nt4}c shows the simulated and measured acoustic mode plots for the in-plane acoustic modes of the $S_2$ set of devices as a function of the bridge size $b$.  As in the main text, the marker size is proportional to the mechanical quality factor. We can easily identify the lower bound of the full bandgap region (gray shaded area) by the abrupt change in the mechanical quality factor and the mechanical \emph{density of states}.  Similar to the $S_1$ devices, as the bridge width is increased, the lower frequency bandgap boundary tunes to higher frequency.  The mechanical mode \emph{localization} is also affected by the presence of the bandgap as seen in Fig.~\ref{Fig_Nt4}d and Fig.~\ref{Fig_Nt4}e. There the displacement field amplitudes of a confined mode are plotted within (Fig.~\ref{Fig_Nt4}d) and on the edge of the bandgap (Fig.~\ref{Fig_Nt4}e).

\section{Mechanical Damping}
\label{appC}
%%%%%%%%%%%%%%%%%%%%%%%%%%%%%%%%%%% FIGURE 03 %%%%%%%%%%%%%%%%%%%%%%%%%%%%%%%%%%%%%%%%%%%%%%%%%%%%%%%%
\begin{figure*}[ht!]
\includegraphics[width=2\columnwidth]{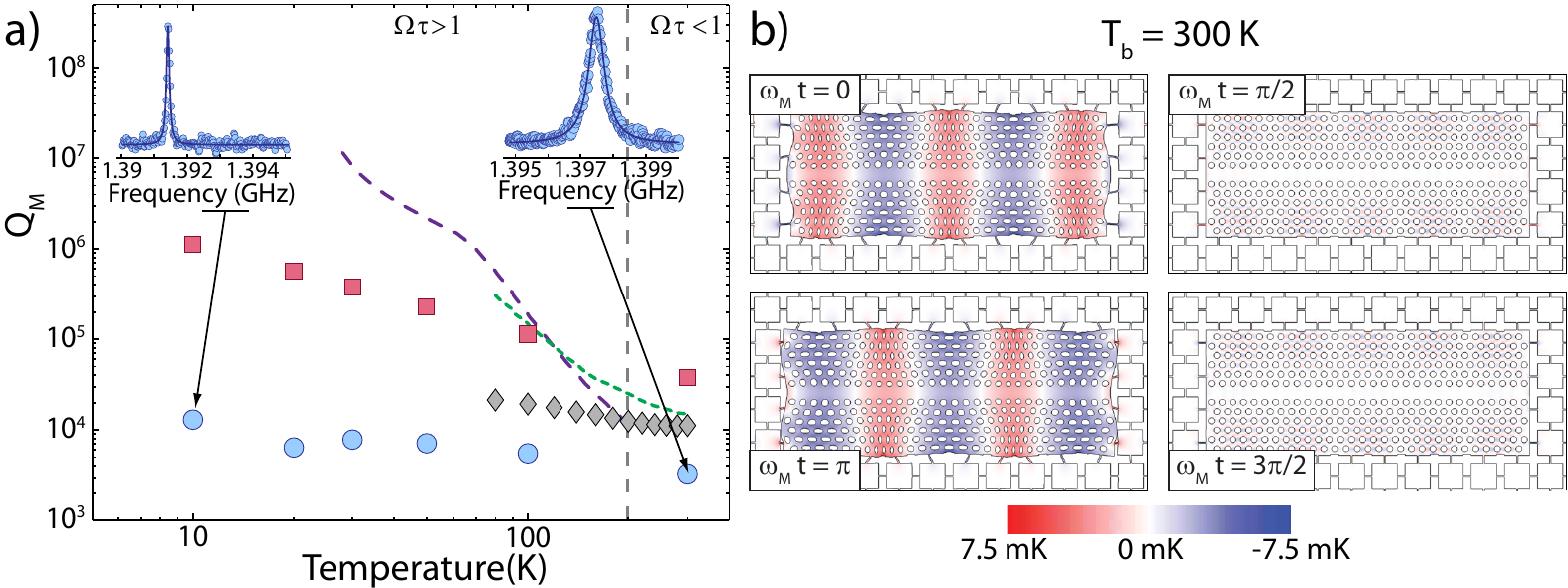} %single column figure - for two-column format
\caption{\label{Fig_TED} \textbf{Temperature-dependent mechanical losses}. \textbf{a}, Comparison between the different sources of mechanical loss with the measured $Q_\text{M}$ values versus temperature. The circles represent the measured values from the $S_1$ samples; diamonds are computed $Q_\text{M}$ from acoustic attenuation data from Ref.~\cite{Att_Si_lambade_1995}; squares represent the simulated values for TED as explained on the text. The purple and green lines are the calculated $Q_\text{M}$ due to Akheiser and Landau-Rumer phonon-phonon dissipation mechanism respectively. The insets show the measured RF PSD at $10$~K and $300$~K for extracting $Q_\text{M}$. \textbf{b} Thermo-mechanical 2D-FEM simulations for the mechanical mode at $1.4$~GHz shown in Fig.~3 of the main text. The thermal profile is plotted at various times during the mechanical cycle.}
\end{figure*}
%%%%%%%%%%%%%%%%%%%%%%%%%%%%%%%%%%%%%%% FIGURE 03 %%%%%%%%%%%%%%%%%%%%%%%%%%%%%%%%%%%%%%%%%%%%%%%%%%%%%%%%

For micro- and nano-mechanical structures the many loss mechanism include clamping losses, dissipation due to phonon-phonon interactions, and surface effects. In the present work, the engineered structures virtually eliminate the clamping loss for modes within the full mechanical bandgap. As a result other loss mechanisms can be studied. In the main text, we present a series of measurement of a confined mode, at $\Omega_\text{M}/2\pi=1.4$~GHz, versus the experimental ambient temperature which shows a temperature dependent mechanical quality factor, $Q_\text{M}$. In Figure~\ref{Fig_TED}(a) we compare our results (blue circles) with other temperature-dependent losses. For the acoustic attenuation, $\alpha(\Omega;T)$,  the relevant bulk measurement results are taken from Ref.~\cite{Att_Si_lambade_1995} and we plot (black diamonds) the calculated $Q_\text{M}=2\pi\Omega_\text{M}/(2\alpha)$, where $\Omega_\text{M}/2\pi=1.4$~GHz is the mechanical frequency. The measured values from Ref.~\cite{Att_Si_lambade_1995} indicate that our devices are not limited by the bulk losses.

In order to provide an upper bound for the quality factor in our structures, we compare our measurements with the temperature-dependent acoustic attenuation provided by the Akheiser~\cite{akhieser_original_1939} (green dashed line), Landau-Rumer~\cite{landau_rumer_absorption_1937} (purple dashed line), and Thermoelastic Damping~\cite{zener_internal_1937,zener_internal_1938, lifshitz_thermoelastic_2000} (TED - red square points) models. The Landau-Rumer model provides a microscopic theory for sound absorption and is valid on the limit of $\Omega_\text{M}\tau\gg1$, where $\tau$ is the mean time between collisions of thermal-phonon in the solid. Akheiser's model treats the dissipation of heat generated by strain trough the Boltzmann transport equation~\cite{woodruff_absorption_sound_1961} and is valid for $\Omega_\text{M}\tau\ll1$. Finally TED considers the heat diffusion in homogeneous materials which can be calculated for any geometry.

The equations for acoustic attenuation for Landau-Rumer ($\alpha_\text{LR}$) and Akheiser's ($\alpha_\text{AK}$) models used are~\cite{woodruff_absorption_sound_1961}:
\begin{equation}
\alpha_\mathrm{LR}(\Omega;T) = \frac{\pi\gamma^2\Omega C_vT}{4\rho c_s^2},\qquad \mathrm{and}\qquad \alpha_\mathrm{AK}(\Omega;T) = \frac{\gamma^2\Omega^2 C_vT\tau}{3\rho c_s^2},
\end{equation}
where $\gamma$ is the average Gr\"{u}neisen coefficient extract from Ref.~\cite{Gruneiser_philip_third-order_1983}, $C_v$ is the volumetric heat capacity with values from Ref.~\cite{Cv_estreicher_thermodynamics_2004}, and $\rho=2330~\text{kg}/\text{m}^3$ and $c_s=9.15\times10^3$~m/s are the density and speed of sound of Si respectively.  Fig.~\ref{Fig_TED}(a) shows that for temperatures above $T=100$~K the Akheiser and Landau-Rumer losses are dominant. Note that $\Omega_\text{M}\tau=1$ for $T\cong200$~K and only Landau-Rumer is valid below this point. In this region TED effects due to the sample geometry become important.

Our approach for calculating TED follows that of Ref.~\cite{duwel_engineering_2006}, where the TED-limited $Q_\text{M,TED}$ is extracted from 2D-FEM simulations~\cite{COMSOL2009} for the thermo-mechanical equations considering a finite thickness. Fig.~\ref{Fig_TED}(b) shows the change in temperature, $\Delta T(\mathbf{r})=T-T_o$, for the deformed structure, versus the phase of the mechanical oscillations, at the expected brownian motion amplitude for $T_o=300$~K. The temperature difference was calculated based on the maximum thermal displacement amplitude $x_\text{max}=\sqrt{2kT_o/ m_\text{eff}\Omega_M^2}$. These plots show that during a mechanical cycle, even when there is no deformation, i.e. $\Omega_\text{M}t=\pi/2$ and $3\pi/2$, the temperature gradient is non-zero.  This indicates that the temperature does not follow adiabatically the strain/stress profile, causing a time-delayed force to be imparted on the resonator, and leading to dissipation.

\end{document}